\documentclass[11pt,a4paper]{article}
\pdfoutput=1

\usepackage{jheppub}
\usepackage{latexsym}
\usepackage{revsymb}
\usepackage{multirow}
\usepackage{color}
\usepackage[usenames,dvipsnames,svgnames,table]{xcolor}

\usepackage{graphicx}
\usepackage{epsfig}  
\usepackage{epsf}    
\usepackage{dcolumn}
\usepackage{bm}
\usepackage{dcolumn}
\usepackage{textcomp}
\usepackage{float}
\usepackage{subfig}
\usepackage{hypcap}
\usepackage[]{hyperref}

\hypersetup{
  bookmarks=true,         
  unicode=false,          
  pdftoolbar=true,        
  pdfmenubar=true,        
  pdffitwindow=true,     
  pdfstartview={FitH},    
  pdfsubject={Neutrino Oscillations Phenomenology},   
  pdfnewwindow=true,      
  pdfcreator={RevTeX},
  colorlinks=true,       
  linkcolor=red,          
  citecolor=blue,        
  filecolor=black,      
  urlcolor=blue,           
}

\def\anu{{\bar\nu}}

\newcommand{\beq}{\begin{equation}}
\newcommand{\eeq}{\end{equation}}
\newcommand{\beqa}{\begin{eqnarray}}
\newcommand{\eeqa}{\end{eqnarray}}

\newcommand{\tx}{{\theta_{12}}}
\newcommand{\ty}{{\theta_{13}}}
\newcommand{\tz}{{\theta_{23}}}

\newcommand{\dl}{{\Delta_{31}}}
\newcommand{\ds}{{\Delta_{21}}}

\newcommand{\atil}{\hat{A}}
\newcommand{\dtil}{\hat{\Delta}}

\newcommand{\dcp}{\delta_{\mathrm{CP}}}
\newcommand{\nova}{NO$\nu$A~}
\newcommand{\pmue}{P(\nu_\mu \rightarrow \nu_e)}

\newcommand{\pmuebar}{P(\bar{\nu}_{\mu} \rightarrow \bar{\nu}_e)}

\newcommand{\dchsq}{\Delta\chi^2}

\begin{document}

\DeclareGraphicsExtensions{.eps,.ps}

\title{The need for an early anti-neutrino run of \nova}

\author[a]{Suprabh Prakash,\footnote{Present address:
Harish-Chandra Research Institute, Chhatnag Road, 
Jhunsi, Allahabad 211019, India}}
\author[a]{Ushak Rahaman,} 
{\author[a]{S. Uma Sankar$\,$}

\affiliation[a]{Department of Physics, Indian Institute of Technology Bombay, Powai, 
Mumbai 400 076, India}

\emailAdd{suprabh@phy.iitb.ac.in}
\emailAdd{ushak@phy.iitb.ac.in}
\emailAdd{uma@phy.iitb.ac.in}

\begin{abstract}
{
The moderately large value of $\ty$, measured recently 
by reactor experiments, is very welcome 
news for the future neutrino experiments. In particular, 
the \nova experiment, with 3 years each of $\nu$ and 
$\anu$ runs, will be able to 
determine the mass hierarchy if one of the following 
two favourable combinations is true: 
normal hierarchy with $-180^\circ \leq \dcp \leq 0$ or 
inverted hierarchy with $0\leq \dcp \leq 180^\circ$.
In this report, we study the hierarchy reach of the
first 3 years of \nova data. 
Since $\sin^2 2 \tz$ is measured to be non-maximal,
$\tz$ can be either in the lower or higher octant.
Pure $\nu$ data is affected by  $\ty$-hierarchy and
octant-hierarchy degeneracies, which limit the hierarchy
sensitivity of such data. A combination of $\nu$ and
$\anu$ data is not subject to these degeneracies and 
hence has much better hierarchy discrimination capability.
We find that, with a 3 year $\nu$ run, hierarchy determination
is possible for only two of the four octant-hierarchy combinations.
Equal 1.5 year runs in $\nu$ and $\anu$ modes give good
hierarchy sensitivity for all the four combinations.
}

\end{abstract}


\maketitle

\section{Introduction}

Neutrino oscillations are one of the most significant 
evidences for physics beyond standard model. The
discovery by the reactor neutrino experiments during the
last two years, that $\ty$ is non-zero, created a lot of
excitement \cite{An:2012eh,Ahn:2012nd,Abe:2012tg}. 
In fact, its measured value is moderately
large and is just below the upper limit established earlier
\cite{Apollonio:1997xe,Apollonio:1999ae,Narayan:1997mk}.
The Daya Bay experiment gives the most precise value:
$\sin^{2}2\ty=0.089\pm 0.01$ \cite{An:2012eh}. 
By the end of Daya Bay's run, the uncertainty 
is expected to be reduced from the present 10$\%$ to 5$\%$
\cite{Kyoto2012DayaBay}. Another important recent discovery 
is the precision measurement of $\sin^{2}2\tz$ 
by MINOS, which found it to be non-maximal \cite{Nichol:2012}. This
raises the problem of determining the true octant of $\tz$. 

Neutrino oscillations depend on two mass-squared differences,
$\ds=m_{2}^{2}-m_{1}^{2}$ and $\dl=m_{3}^{2}-m_{1}^{2}$, 
three mixing angles and a CP violating phase $\dcp$. Here 
$m_{1}$, $m_{2}$ and $m_{3}$ are the masses of three mass eigenstates. 
The present oscillation data determine the mass-squared differences and
mixing angles reasonably well 
\cite{Tortola:2012te,Fogli:2012ua,GonzalezGarcia:2012sz}. 
The observed energy dependence of the solar neutrino  
survival probability requires $\ds$ to be positive. But the
present data allow $\dl$ to be either positive or negative.
The case of positive $\dl$ is called normal hierarchy (NH)
and that of negative $\dl$ is called inverted hierarchy (IH).
If the lightest neutrino mass is negligibly small, we have
the following patterns: $m_{3}\gg m_{2}>m_{1}$ for NH and 
$m_{2}>m_{1}\gg m_{3}$ for IH. It is possible that all 
the three masses are nearly degenerate. In such a situation also
the data allows either hierarchy.
Determination of the neutrino mass hierarchy, the octant of $\tz$ 
and the search for CP violation in neutrino sector are the important 
physics goals of current and future oscillation experiments. 

A number of models are proposed to explain the observed 
pattern of neutrino masses and mixing. Among these,
the models predicting NH are qualitatively different
from those predicting IH. Therefore, the determination of the 
neutrino mass hierarchy will enable us to distinguish between
different types of models \cite{Albright:2004kb}. 
A large number of these models 
predict $\ty$ to be zero and $\tz$ to be maximal. A precise
measurement of the deviations from these predictions will
enable us to discern the pattern of symmetry breaking in 
the models. Ever since the possibility of generating baryon asymmetry
via leptogenesis was raised \cite{Fukugita:1986hr},
the search for leptonic CP violation has acquired great 
significance. 

A simple way to achieve the above three goals is to measure 
the probabilities for $\nu_{\mu}\rightarrow \nu_{e}$ oscillation 
($\pmue$) and $\bar{\nu}_{\mu}\rightarrow \bar{\nu}_{e}$ 
oscillation ($\pmuebar$). The leading term in both these probabilities 
is proportional to $\sin^{2}2\ty \sin^{2}\tz$. Therefore, 
the moderately large value of $\ty$ makes it possible  
for the current experiments to address the problems of 
both hierarchy and the octant of $\tz$. Appreciable matter effects
in the \nova experiment make it an excellent tool to determine
the hierarchy for favourable values of parameters 
\cite{Huber:2009cw,Prakash:2012az}. In addition, T2K and \nova 
can determine octant of $\tz$ at 2$\sigma$ 
\cite{Agarwalla:2013ju,Chatterjee:2013qus}
for all values of $\dcp$.  

\section{Degeneracies in $\pmue$ and $\pmuebar$}

Among the neutrino oscillation parameters, there are two small
quantities: $\ty$ and $\alpha = \ds/\dl$. By setting one or both
to be zero, it was possible so far, to reduce all the measured 
survival probabilities to effective two flavour formulae. In the
$\nu_e$ appearance measurements at T2K and NO$\nu$A, the first 
non-trivial three flavour oscillation effects will be observed,
which are proportional to the small quantities $\ty$ and $\alpha$.
In the approximation of keeping only the terms which are second
order in these small quantities, the $\nu_\mu \rightarrow \nu_e$ 
oscillation probability is given by \cite{Cervera:2000kp,Freund:2001pn},
\begin{eqnarray}
\pmue & = & 
\sin^2 2 \ty \sin^2 \tz\frac{\sin^2\dtil(1-\atil)}{(1-\atil)^2} \nonumber\\ 
& & +\alpha \cos \ty \sin2\tx \sin 2\ty \sin 2\tz \cos(\dtil+\dcp)
\frac{\sin\dtil \atil}{\atil} \frac{\sin \dtil(1-\atil)}{1-\atil} \nonumber\\
 & & +\alpha^2 \sin^2 2 \tx \cos^2 \ty \cos^2 \tz 
 \frac{\sin^2 \dtil \atil}{\atil^2}.
\label{pmue-exp}
\end{eqnarray}
Here $\dtil = \dl L/4 E$ and $\atil = A/\dl$, where $A$ 
is the Wolfenstein matter term \cite{msw1}. The expression for 
$\pmuebar$ is obtained by changing the signs of $\atil$ and $\dcp$
in $\pmue$. $\dl$ is positive for NH and is negative for IH. 
From Eq.~(\ref{pmue-exp}), we see 
that the oscillation probability depends on unknowns, i.e. hierarchy, 
octant of $\tz$ and $\dcp$, along with other parameters, such as $\ty$. A measurement 
of these probabilities, in general, gives rise to degenerate solutions.

\subsection{Hierarchy-$\dcp$ degeneracy}

From the current measurements, we know that $\sin 2 \ty \approx 0.3$
whereas $|\alpha| \approx 0.03$. Hence, the first term in $\pmue$
(and in $\pmuebar$) is much larger than second term and the third
term is completely negligible. The largest amount of matter effect
and hence hierarchy sensitivity, comes from the leading term. 
For NH (IH), the first term in $\pmue$ becomes larger (smaller). 
For $\pmuebar$, the situation is reverse. These changes in $\pmue$
and in $\pmuebar$ can be {\it amplified or canceled} by the second 
term, depending on the value of $\dcp$.  This is illustrated in 
Fig. \ref{prob}, where $\pmue$ and $\pmuebar$ are plotted for 
the \nova experiment. For NH and $\dcp$ in the lower half plane 
(LHP) ($-180^{\circ}\leq\dcp\leq0$),  the values of $\pmue$ 
($\pmuebar$) are reasonably greater (lower) than the values of 
$\pmue$ ($\pmuebar$) for IH and any value of $\dcp$. Similarly,  
for IH and $\dcp$ in the upper half plane (UHP)
($0\leq\dcp\leq180^{\circ}$) the values of $\pmue$ ($\pmuebar$) 
are reasonably lower (greater) than the values of 
$\pmue$ ($\pmuebar$) for NH and any value of $\dcp$. 
Hence, for these {\it favourable combinations}, \nova is capable 
of determining the hierarchy at a confidence level (C.L.) of 
$2\sigma$ or better, with 3 years each of $\nu$ and $\anu$ runs. 
However, as mentioned above, the change in the first term can be 
canceled by the second term for unfavourable values of
$\dcp$. This leads to hierarchy-$\dcp$ 
degeneracy \cite{Barger:2001yr, Minakata:2003wq, Mena:2004sa}. 
From Fig.~(\ref{prob}), we see that,   
$\pmue$ and $\pmuebar$ for NH and $\dcp$ in the UHP 
are very close to or degenerate with 
those of IH and $\dcp$ in the LHP.
For these {\it unfavourable combinations},
\nova has {\bf no} hierarchy sensitivity \cite{Prakash:2012az}. 
Addition of T2K data gives rise 
to a small sensitivity \cite{Mena:2004sa,Agarwalla:2012bv}.
In this paper, we explore the further degeneracies in the 
case of the favourable hierarchy-$\dcp$ combinations.

\begin{figure}[H]

\centering
\includegraphics[width=0.8\textwidth]{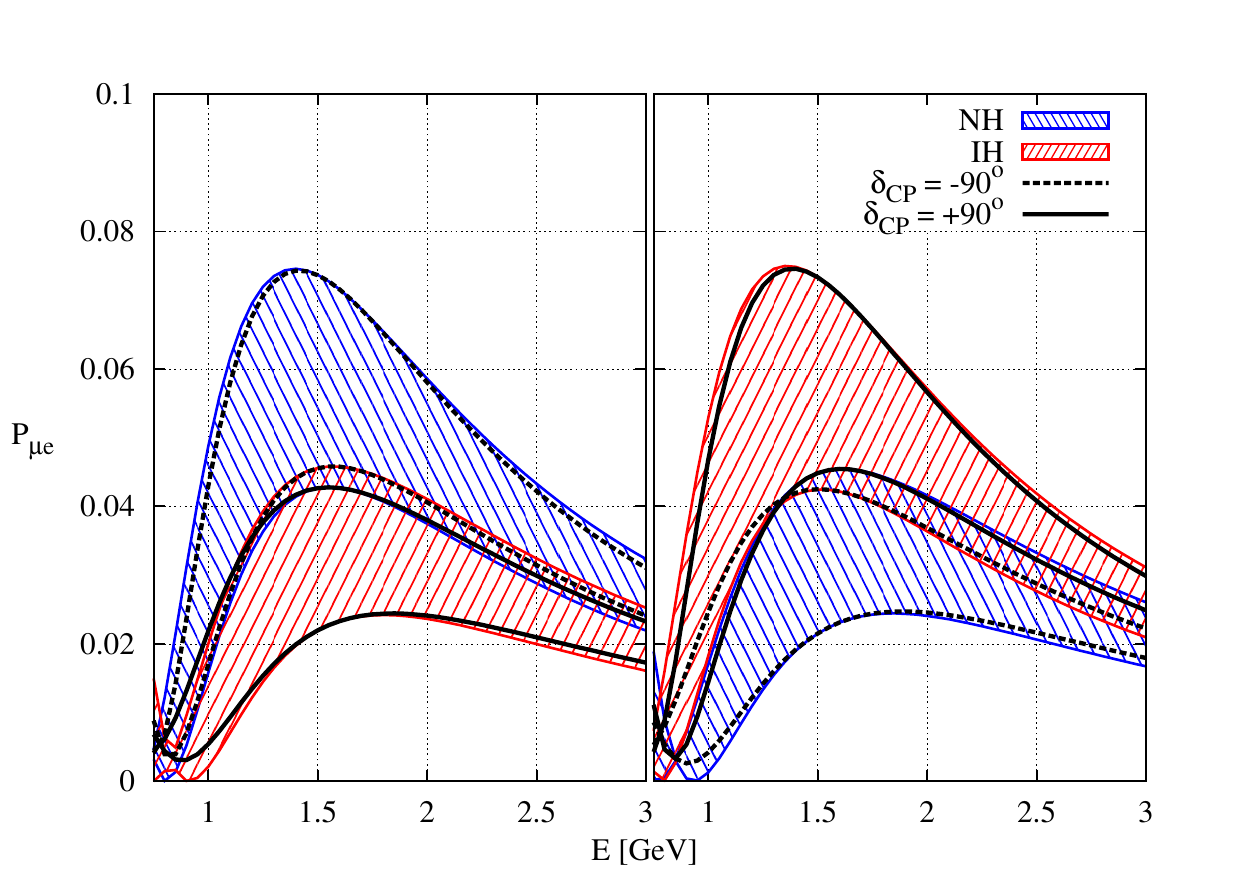}
\caption{\footnotesize{$\pmue$ (left panel) and $\pmuebar$ (right panel) vs. energy for
NO$\nu$A. Variation of $\dcp$ 
leads to the blue (red) bands for NH (IH). 
The plots are drawn for
maximal $\tz$ and other neutrino parameters given in the text.}}
\label{prob}
\end{figure}

\subsection{$\ty$-hierarchy degeneracy}

Even if $\dcp$ is in the favourable half-plane, there are 
further degeneracies which limit the hierarchy sensitivity 
of an experiment. For example, in Eq.~(\ref{pmue-exp}), the 
increase (reduction) in the first term for NH (IH) case,
due to matter effect, can 
be canceled by choosing a lower (higher) value of $\ty$. 
This $\ty$-hierarchy degeneracy \cite{Barger:2001yr} can 
reduce the hierarchy sensitivity. However, a combination of
$\nu$ and $\anu$ data is not susceptible to this degeneracy.
The reason is the following. In $\nu$ data, it is possible 
to have $\pmue(\ty, NH) \approx \pmue(\ty^\prime, IH)$ with
$\ty^\prime > \ty$. However, for such a choice of $\ty^\prime$,
we will have $\pmuebar(\ty, NH)$ significantly smaller than 
$\pmuebar(\ty^\prime, IH)$. Thus a degeneracy in the $\nu$ data
is resolved by the $\anu$ data (and vice-verse). If the allowed
range of $\ty$ is large, then a combination of $\nu$ and $\anu$
data has better hierarchy sensitivity compared to pure $\nu$ 
data. 

\subsection{Octant-hierarchy degeneracy}

A more serious degeneracy, which limits the hierarchy sensitivity,
is the octant-hierarchy degeneracy. MINOS experiment has 
measured $\sin^2 2 \tz < 1$ \cite{Kyoto2012MINOS} and the
global fits favour a non-maximal value of $\tz$ 
\cite{Tortola:2012te,Fogli:2012ua,GonzalezGarcia:2012sz}. 
There are two degenerate solutions, with $\tz$
in the lower octant (LO) ($\sin^2 \tz < 0.5$) and with
$\tz$ in the higher octant (HO) ($\sin^2 \tz > 0.5$).
Thus we have four possible octant-hierarchy combinations:
LO-NH, HO-NH, LO-IH and HO-IH. As already stated, the 
first term in $\pmue$ becomes larger (smaller) for NH (IH).
The same term also becomes smaller (larger) for LO (HO).
If the case HO-NH (LO-IH) is true, then the values of $\pmue$ 
are significantly higher (smaller) than those for IH (NH) and 
any octant. For these two cases, pure $\nu$ data has good
hierarchy determination capability. But the situation is 
very different for the two cases LO-NH and HO-IH.
The increase (decrease) in the first term of $\pmue$ due to NH 
(IH) is canceled (compensated) by the choice of LO (HO).
Thus the two cases, LO-NH and HO-IH, have degenerate values 
for $\pmue$. However, this degeneracy is not present in
$\pmuebar$, which receives a double boost (suppression) for the 
case of HO-IH (LO-NH). Thus the octant-hierarchy degeneracy in 
$\pmue$ is broken by $\pmuebar$ (and vice-verse) as in the case
of $\ty$-hierarchy degeneracy. Therefore pure $\nu$ data has 
{\bf no} hierarchy sensitivity if the cases LO-NH or HO-IH
are true, but a combination of $\nu$ and $\anu$ data will
have a good sensitivity.

\section{Results}

\subsection{Simulation Details}

In this report, we study the possible hierarchy
reach of the first three years of \nova data. As shown in 
the previous section, a pure $\nu$ data is subject to 
$\ty$-hierarchy and octant-hierarchy degeneracies, whereas
a combination of $\nu$ and $\anu$ data is not. Therefore,
here we consider two options: (a) a 3 year $\nu$ run 
(labeled 3$\nu$ in the rest of the paper) and (b) equal
$\nu$ and $\anu$ runs of 1.5 years each (labeled 1.5$\nu$+1.5$\anu$). 

\nova experiment \cite{nova_tdr} consists of a 14 kiloton totally 
active scintillator detector (TASD),  
placed 810 km away from Fermilab, situated at a 
$0.8^\circ$ off-axis location from the NuMI beam. 
The $\nu$ flux peaks sharply at 2 GeV,
close to the energy range 1.4-1.8 GeV, where the
oscillation maxima occur for NH and for IH. It is scheduled
to have equal $\nu$ and $\anu$ runs of 3 years each, with a NuMI
beam power of 700 kW, corresponding to $6 \times 10^{20}$ protons 
on target per year. In our simulations, we have used the re-tuned
signal acceptance and background rejection factors taken from 
\cite{Kyoto2012nova,Agarwalla:2012bv}. In the numerical simulations, 
we took the solar oscillation parameters to be
$\sin^2 \tx = 0.30$ and $\ds = 7.5 \times 10^{-5}$ eV$^2$,
which have been kept fixed \cite{GonzalezGarcia:2012sz}. 
The other parameters used are 
$\sin^2 2 \ty = 0.089$ and $\Delta m^2_{\rm eff} = \pm 2.4 \times
10^{-3}$ eV$^2$ \cite{Nichol:2012}, where the positive (negative) 
sign is for NH (IH).
$\dl$ is derived from $\Delta m^2_{\rm eff}$ from the expression
given in \cite{Nunokawa:2005nx}. For $\tz$, we considered the 
cases of both maximal and non-maximal mixing. For maximal mixing (MM),
$\sin^2 \tz = 0.5$. For non-maximal mixing, we have used 
the two degenerate best-fit values of the global fits: 0.41 for
$\tz$ in LO and 0.59 for $\tz$ in HO \cite{GonzalezGarcia:2012sz}. 

The spectrum of electron neutrino appearance events 
and that of the electron anti-neutrino appearance events 
are first computed for an assumed true hierarchy. The same quantities
are calculated again for the wrong hierarchy and the $\dchsq$ 
is computed between the event spectra for the true and the 
wrong hierarchies. The event spectrum simulations and the 
$\dchsq$ calculation are done by using the software GLoBES 
\cite{Huber:2004ka,Huber:2007ji}. The minimum $\dchsq$ is 
computed by doing a marginalization over the neutrino parameters. 
We took $\sigma (\Delta m^2_{\rm eff}) =3 \%$ \cite{Itow:2001ee} 
and $\sigma (\sin^2 2 \ty) = 10\%$ in the preliminary 
calculations and $5\%$ in later calculations. For both these
parameters, the marginalization was done over $2 \sigma$ range 
with Gaussian priors.
The marginalization range for $\sin^2 \tz$ is its
$3 \sigma$ allowed range: $[0.35, 0.65]$ 
and that of $\dcp$ is the full range $[-180^\circ, 180^\circ]$
\footnote{The global best fit \cite{nufit,Capozzi:2013csa}
indicates a preference for $\dcp$ to be in the LHP. But here,
we will be conservative and consider the full range of $\dcp$
in our marginalization.}. No priors were added for these two parameters.

\subsection{Effect of precision of $\sin^22\ty$ on hierarchy determination}

\begin{figure}[H]
\centering
\includegraphics[width=0.8\textwidth]{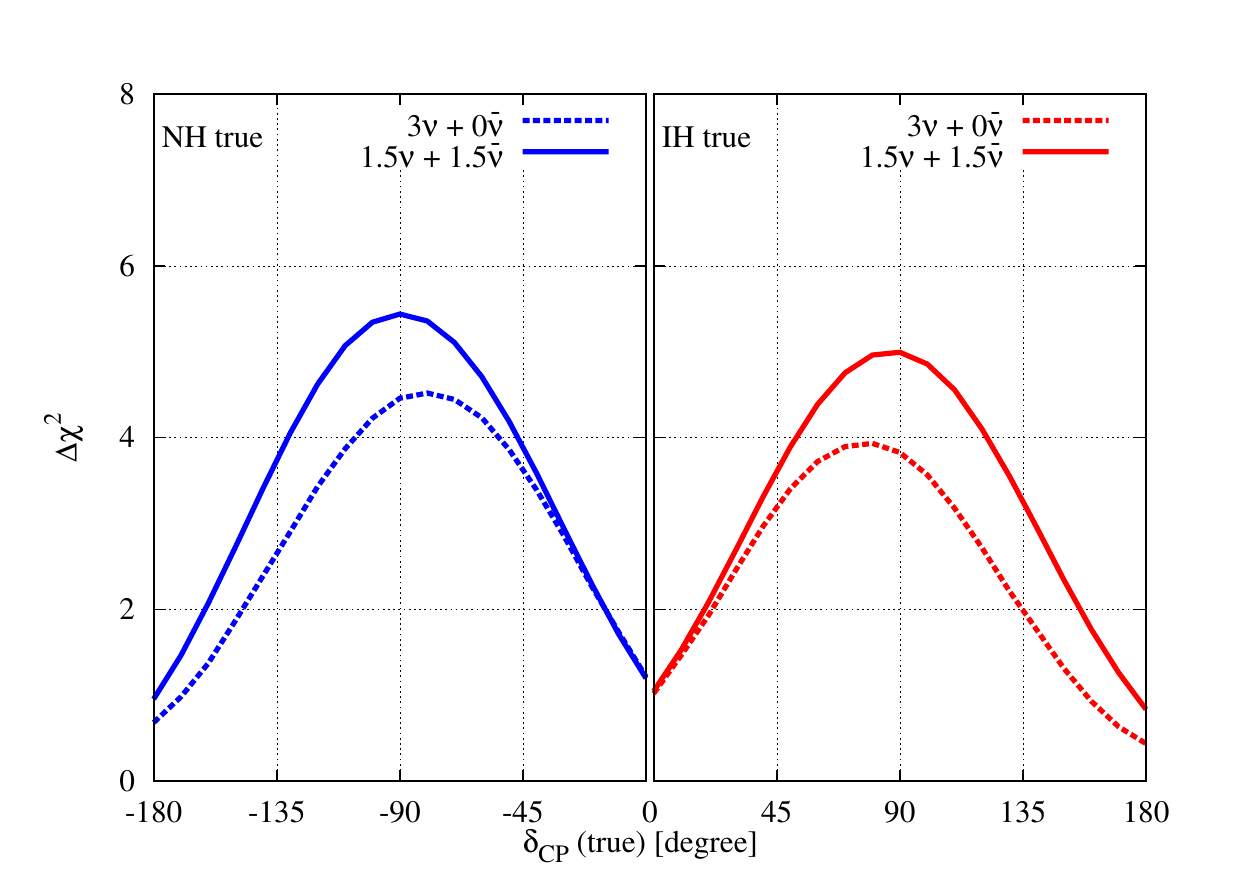}
\caption{\footnotesize{Hierarchy sensitivity assuming 10$\%$ uncertainty in 
$\sin^22\theta_{13}$ and maximal $\theta_{23}$. 
In the left (right) panel, the true hierarchy is taken to be
NH (IH).}}
\label{tenpc}
\end{figure} 
In Fig. \ref{tenpc} we have shown the hierarchy determination 
potential of \nova assuming a 10$\%$ uncertainty in $\sin^22\ty$. The plots 
show $\dchsq$ vs. $\dcp$(true) for $\tz = 45^\circ$, for both 
3$\nu$ and $1.5\nu+1.5\anu$ runs. 
The left panel is for NH and LHP and the right panel is 
for IH and UHP. We see from these plots 
that a 2$\sigma$ hierarchy determination is possible 
for about 50$\%$ of the favourable half plane for $1.5\nu+1.5\anu$ run, 
whereas a 3$\nu$ run can determine hierarchy 
for only a smaller range. In particular, 
if IH and UHP is true, a 2$\sigma$ hierarchy determination 
is not possible for any $\dcp$. 
Here the number of $\sigma$ is taken to be $\sqrt{\dchsq}$. 
The lower sensitivity of 3$\nu$ run is due to the marginalization 
over $\ty$. Because of the relatively large range of variation 
for $\ty^\prime$, it is possible for $\pmue(\ty^\prime,IH)$ to
come reasonably close to $\pmue(\ty, NH)$, thus reducing the 
$\dchsq$. As explained in the previous section, the $1.5\nu+1.5\anu$ 
run is less sensitive to this marginalization and gives a larger
$\dchsq$.
If the uncertainty in $\sin^2 2 \ty$ is reduced to $5\%$.
the hierarchy reach for 3$\nu$ does improve and becomes 
equal to that of $1.5\nu+1.5\anu$ run. 

\subsection{Resolving the octant-hierarchy degeneracy}

We now assume that $\sigma(\sin^2 2 \ty) = 5\%$ and 
take $\tz$ to be non-maximal. Once again we limit ourselves to
the favourable hierarchy-$\dcp$ combinations, NH and LHP and 
IH and UHP. But, because of the octant degeneracy of $\tz$, 
we must consider four possible combinations of octant and hierarchy: 
LO-NH, HO-NH, LO-IH and HO-IH. 

\begin{figure}[H]
\centering
\includegraphics[width=0.8\textwidth]{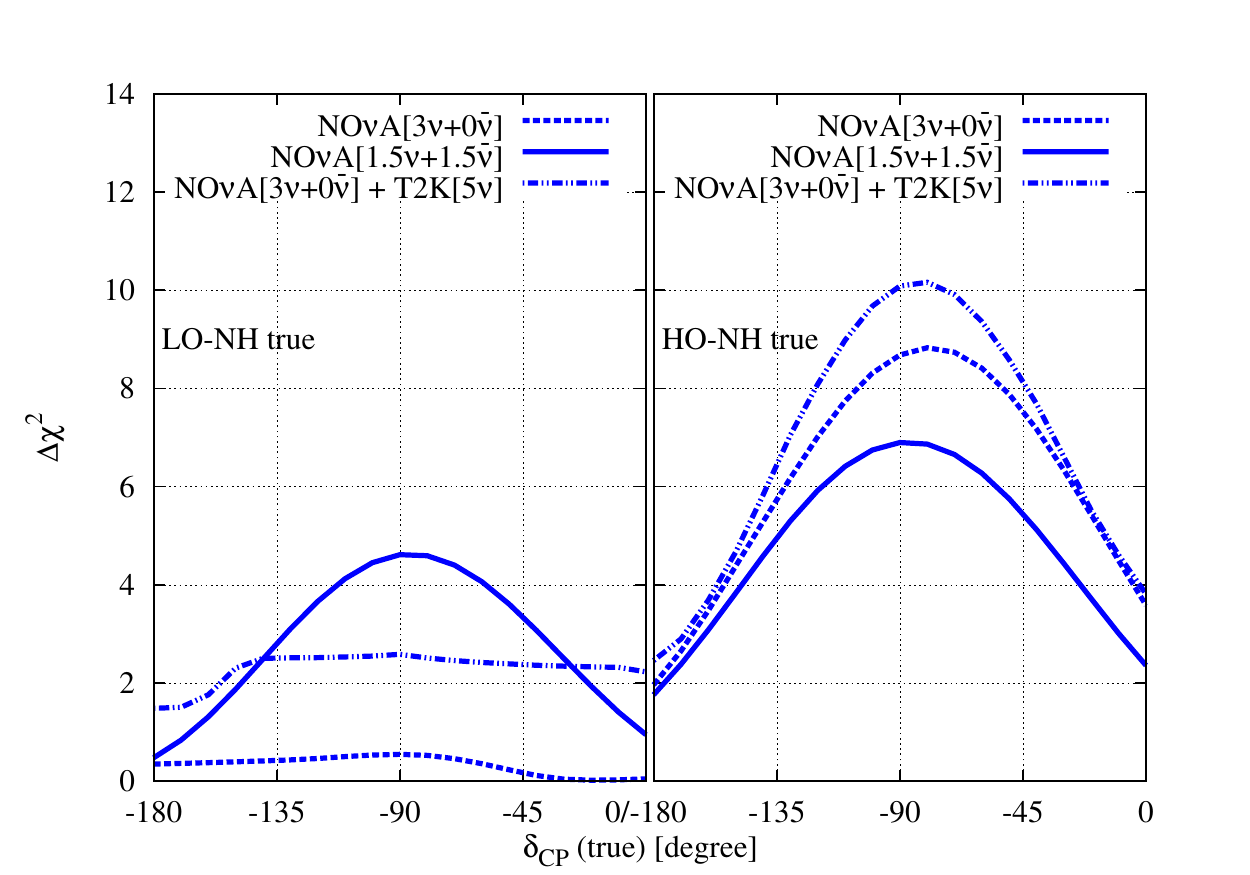}
\caption{\footnotesize{Hierarchy sensitivity assuming 5$\%$ uncertainty in 
$\sin^22\theta_{13}$ for NH and LHP. In the left (right) panel, 
the true $\sin^2\tz$ is taken to be 0.41 (0.59).}}
\label{NH5pc}
\end{figure}

In Fig. \ref{NH5pc}, we show the hierarchy capability assuming 
NH and LHP. The left (right) panel corresponds to $\tz$ in LO (HO). 
In Fig. \ref{IH5pc}, we do the same for IH and UHP. 
From these figures, we see that for HO-NH and LO-IH,
$3\nu$ run does have a better hierarchy reach compared to
$1.5\nu+1.5\anu$ run and is capable of giving a better than
$2 \sigma$ hierarchy discrimination for more than half of the
favourable half plane. But, for the other two possibilities,
LO-NH and HO-IH, $3\nu$ run has {\bf no} hierarchy sensitivity 
whereas $1.5\nu+1.5\anu$ run has reasonable hierarchy sensitivity. 
The very small values of $\dchsq$, for the $3\nu$ run, 
occur due to the marginalization over $\sin^2 \tz$ and $\dcp$. 
Addition of 5 year $\nu$ data from T2K leads only to a small improvement.

As mentioned before,
the dominant term in $\pmue$ is proportional to $\sin^2 
2 \ty \sin^2 \tz$. Matter effects in NH make this term
larger and choosing HO makes it even larger. Hence, for
$\dcp$ in LHP, $\pmue$(HO-NH) is significantly higher than 
$\pmue$(IH) for any values of neutrino parameters. Because 
of the double increase in the probability, the statistics 
for HO-NH will be quite large. Hence, this combination has 
$2 \sigma$ hierarchy discrimination for $87\%$ ($68\%$) 
of the favourable half-plane for $3\nu$ ($1.5\nu+1.5\anu$) run.
Matter effects in IH make the leading term in $\pmue$ smaller 
and choosing LO makes it even smaller. So, for $\dcp$ in UHP, 
$\pmue$(LO-IH) is significantly smaller than $\pmue$(NH)
for any values of neutrino parameters. This double decrease
in probability, leads to the lowest statistics for LO-IH.
Here, $3\nu$ ($1.5\nu+1.5\anu$) run can determine hierarchy
at $2 \sigma$ for $35 \%$ ($20\%$) of favourable half-plane.
However, it must be emphasized that, in these two cases
HO-NH and LO-IH, the hierarchy reach of $1.5\nu+1.5\anu$ is
only slightly worse than that of $3\nu$.

\begin{figure}[H]
\centering
\includegraphics[width=0.8\textwidth]{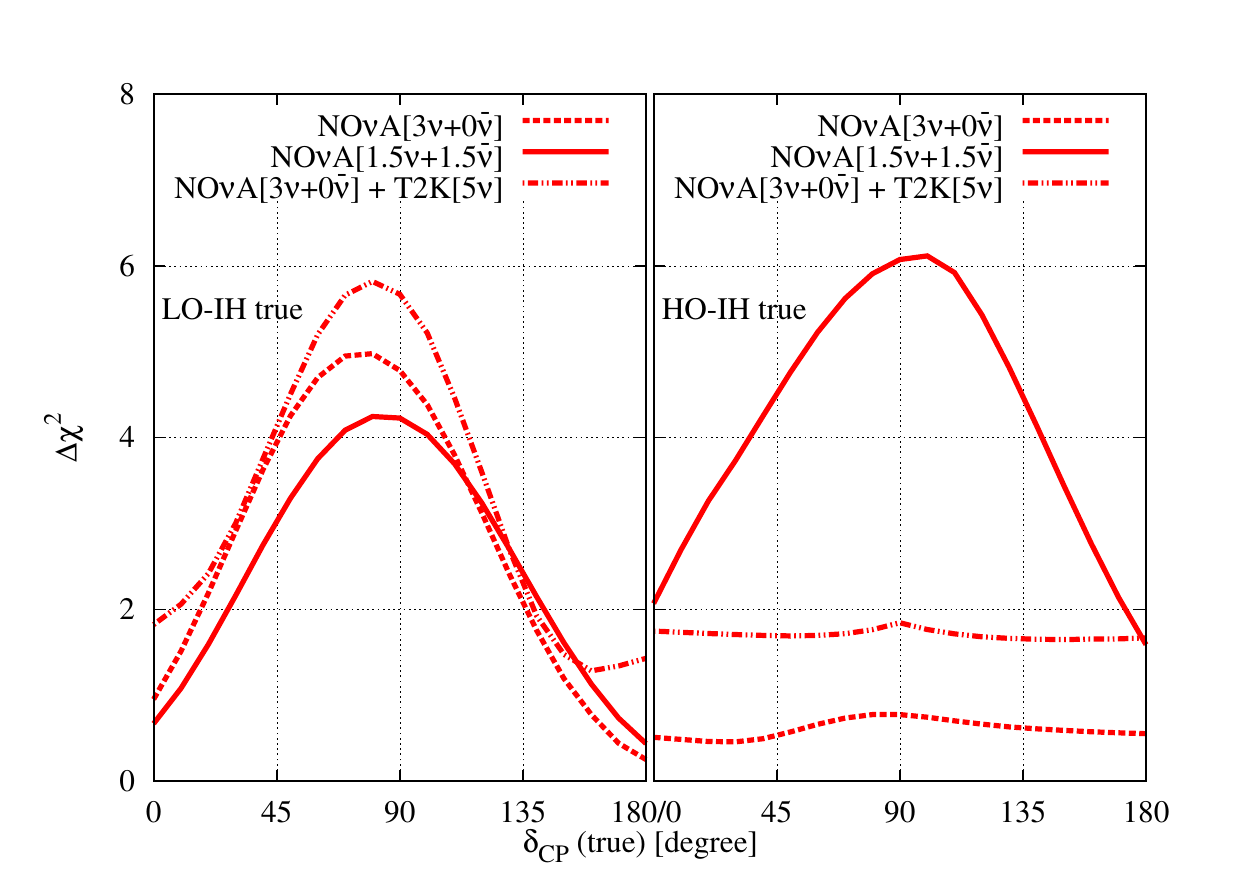}
\caption{\footnotesize{Hierarchy sensitivity assuming 5$\%$ uncertainty in 
$\sin^22\theta_{13}$ for IH and UHP. In the left (right) panel, 
the true $\sin^2\tz$ is taken to be 0.41 (0.59).}}
\label{IH5pc}
\end{figure}

But, for the combination of LO-NH, the choice of NH 
increases $\pmue$ whereas the choice of LO lowers it.
Similarly, for the combination HO-IH, the choice of
IH lowers $\pmue$ and the choice of HO increases it.
The marginalization over $\tz$ and $\dcp$ leads
to a wrong hierarchy probability being very close
to the true hierarchy probability.
Thus, it is possible to have $\pmue$(NH, $\tz < 45^\circ$, $\dcp$) 
mimic $\pmue$(IH, $\tz^\prime > 45^\circ$, $\dcp^\prime)$, where
$\tz$ and $\tz^\prime$ may or may not be complementary and $\dcp$
and $\dcp^\prime$ may or may not be equal.  
But, in the case of $\anu$, 
both the choices LO and NH lead to a reduction in 
the probability and both the choices HO and IH increase 
the probability. Whenever it is possible to have 
$\pmue$(NH, $\tz$, $\dcp$) $\approx\pmue$(IH, $\tz^\prime$, $\dcp^\prime$), 
the corresponding values of $\pmuebar$ will be far apart. 
This is illustrated in Fig. \ref{deg-prob} for two cases,
where $\tz$ and $\tz^\prime$ are complementary. 
For the two left panels $\dcp = \dcp^\prime$ and
for the two right panels $\dcp \neq \dcp^\prime$. 
The large separation in $\pmuebar$ leads to a far better hierarchy 
discrimination for $1.5\nu+1.5\anu$ run compared to $3\nu$ run. 
All the results discussed above are neatly summarized in the
table I. 
In all cases, the $1.5\nu+1.5\anu$ data is insensitive
to the uncertainty in $\sin^2 2 \ty$. Except for the no-sensitivity
combinations, LO-NH and HO-IH,
the $3\nu$ data shows noticeable 
improvement when the uncertainty is reduced to $5\%$ but none
with further reduction to $2\%$. 

\begin{figure}[H]
\centering
\includegraphics[width=0.8\textwidth]{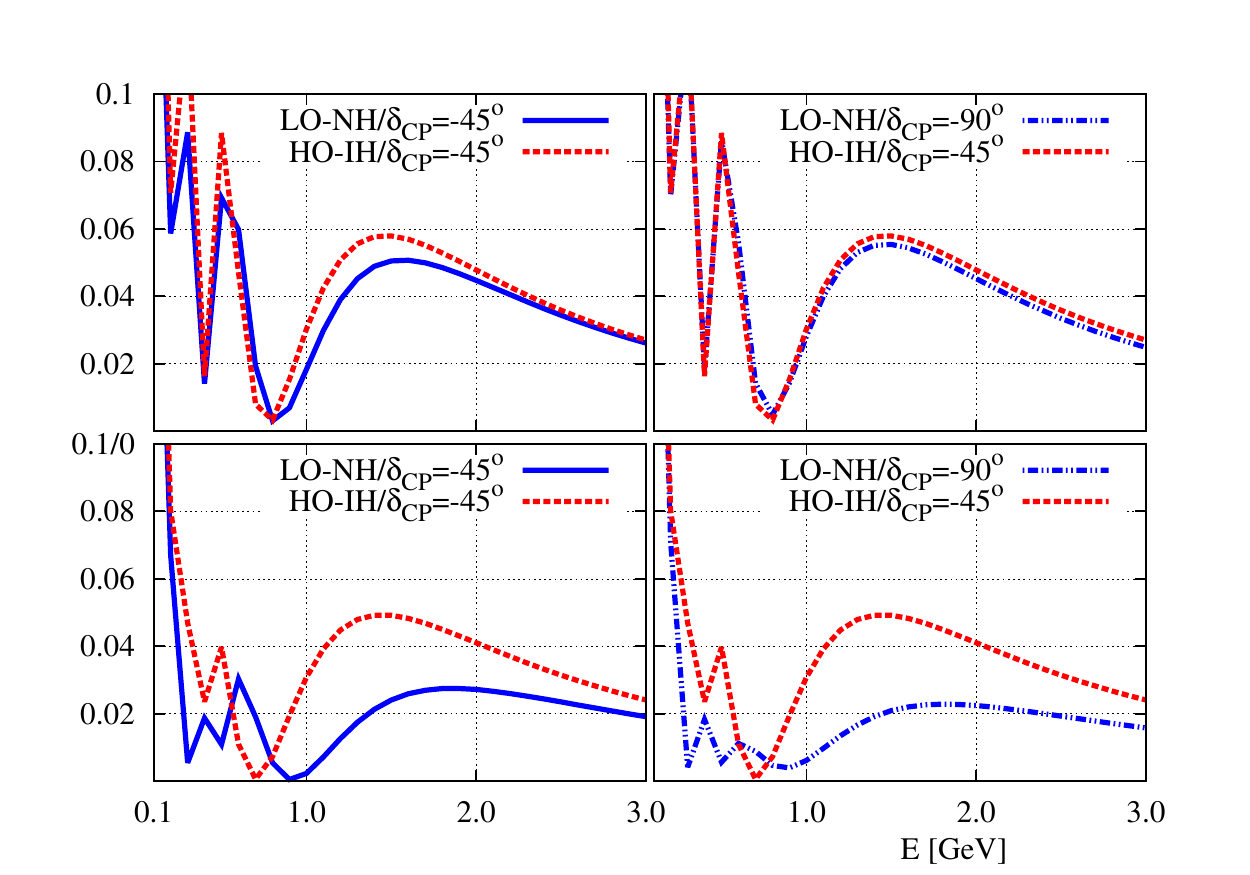}
\caption{\footnotesize{Illustration of degenerate $\pmue$ and non-degenerate $\pmuebar$ for the following two cases.
Left: (LO-NH, $\dcp=-45^\circ$) and 
(HO-IH, $\dcp^\prime=-45^\circ$), 
Right: (LO-NH, $\dcp=-90^\circ$) and 
(HO-IH, $\dcp^\prime=-45^\circ$).}}
\label{deg-prob}
\end{figure}

\begin{table}[H]
\centering
{\footnotesize
\begin{tabular}{||c||ccc||ccc||ccc||}
 \hline
 \hline
\multicolumn{1}{||c||}{$\dcp$:}
&\multicolumn{3}{c||}{$-180^\circ$}
&\multicolumn{3}{c||}{$-135^\circ$}
&\multicolumn{3}{c||}{$-90^\circ$}\\
\cline{2-10}
\multicolumn{1}{||c||}{$\sin^2\tz$:}
&\multicolumn{1}{c}{0.41}
&\multicolumn{1}{c}{0.5}
&\multicolumn{1}{c||}{0.59}

&\multicolumn{1}{c}{0.41}
&\multicolumn{1}{c}{0.5}
&\multicolumn{1}{c||}{0.59}

&\multicolumn{1}{c}{0.41}
&\multicolumn{1}{c}{0.5}
&\multicolumn{1}{c||}{0.59}\\
\hline
\hline

\multicolumn{1}{||c||}{}
& 0.47 & 0.97 & 1.76
& 2.80 & 3.76 & 4.91 
& 4.52 & 5.52 & 6.71 
\\
$1.5\nu+1.5\anu$
& 0.47 & 0.97 & 1.76 
& 2.80 & 3.76 & 4.95 
& 4.61 & 5.53 & 6.93 
\\

& 0.47 & 0.97 & 1.76 
& 2.80 & 3.76 & 4.95 
& 4.61 & 5.53 & 6.96 
\\
\hline
\hline

\multicolumn{1}{||c||}{}
& 0.56 & 0.75 & 1.66 
& 1.10 & 2.90 & 4.61 
& 1.23 & 4.65 & 6.89 
\\
$3\nu+0\anu$
& 0.56 & 0.75 & 1.98 
& 1.10 & 3.37 & 5.76 
& 1.23 & 5.65 & 8.68 
\\

& 0.56 & 0.75 & 2.10 
& 1.10 & 3.61 & 6.21 
& 1.23 & 6.06 & 9.45 
\\
\hline
\hline
\end{tabular}
}
 

{\footnotesize
\begin{tabular}{||c||ccc||ccc||ccc||ccc||ccc||}
 \hline
 \hline
\multicolumn{1}{||c||}{$\dcp$:}
&\multicolumn{3}{c||}{$0$}
&\multicolumn{3}{c||}{$45^\circ$}
&\multicolumn{3}{c||}{$90^\circ$}
\\
\cline{2-10}
\multicolumn{1}{||c||}{$\sin^2\tz$:}
&\multicolumn{1}{c}{0.41}
&\multicolumn{1}{c}{0.5}
&\multicolumn{1}{c||}{0.59}

&\multicolumn{1}{c}{0.41}
&\multicolumn{1}{c}{0.5}
&\multicolumn{1}{c||}{0.59}

&\multicolumn{1}{c}{0.41}
&\multicolumn{1}{c}{0.5}
&\multicolumn{1}{c||}{0.59}
\\
\hline
\hline

\multicolumn{1}{||c||}{}
& 0.67 & 1.10 & 1.92 
& 2.91 & 3.70 & 4.37 
& 4.17 & 5.08 & 5.89 
\\
$1.5\nu+1.5\anu$
& 0.67 & 1.10 & 2.10 
& 3.03 & 3.81 & 4.53 
& 4.22 & 5.16 & 6.11 
\\

& 0.67 & 1.10 & 2.10 
& 3.03 & 3.81 & 4.53
& 4.22 & 5.16 & 6.12 
\\
\hline
\hline

\multicolumn{1}{||c||}{}
& 0.74 & 1.07 & 0.51 
& 3.10 & 3.20 & 0.53 
& 3.78 & 3.81 & 0.77 
\\
$3\nu+0\anu$
& 0.99 & 1.41 & 0.51 
& 3.92 & 4.05 & 0.53 
& 4.77 & 4.83 & 0.77 
\\

& 1.02 & 1.52 & 0.51 
& 4.28 & 4.43 & 0.53 
& 5.16 & 5.23 & 0.77 
\\
\hline
\hline
\end{tabular}
}

\caption{\footnotesize{Hierarchy discrimination reach of \nova data for
$1.5\nu+1.5\anu$ and $3\nu$ runs. The upper (lower) half
is for NH-LHP (IH-UHP) true. In each case, the $\dchsq$ values are
shown for $\tz$ being in LO, maximal and in HO and for 
three values of $\dcp$, covering half of the favourable half
plane. $\dchsq$ values for the other half are nearly symmetric 
about $\dcp = \pm 90^\circ$, as can be seen from the figures. 
\ref{NH5pc} and \ref{IH5pc}. The three lines in each small box
correspond to $10\%$, $5\%$ and $2\%$ precision in $\sin^2
2 \ty$ respectively.}}

\end{table}


In the most recent global fits of the neutrino oscillation data
\cite{nufit}, the best-fit value of $\sin^2 \tz$ in LO is 
$0.45$, ({\it i. e.} closer to the maximal mixing value), though 
the best-fit value in HO remains at $0.59$. We have redone our 
calculations and compared the hierarchy discrimination ability 
of $3 \nu$ vs $1.5 \nu + 1.5 \anu$ data of NO$\nu$A, for these
new values of $\sin^2 \tz$. These results are shown in 
Figs.~\ref{newtzNH} and~\ref{newtzIH}. As we see from these 
figures, even with the smaller deviation of $\tz$ from maximality,
the $3 \nu$ run of \nova has {\bf no} hierarchy sensitivity 
for the two combinations LO-NH and HO-IH, whereas the $1.5\nu +
1.5\anu$ run has good hierarchy determination capability for all
four combinations.

\begin{figure}[H]
\centering
\includegraphics[width=0.8\textwidth]{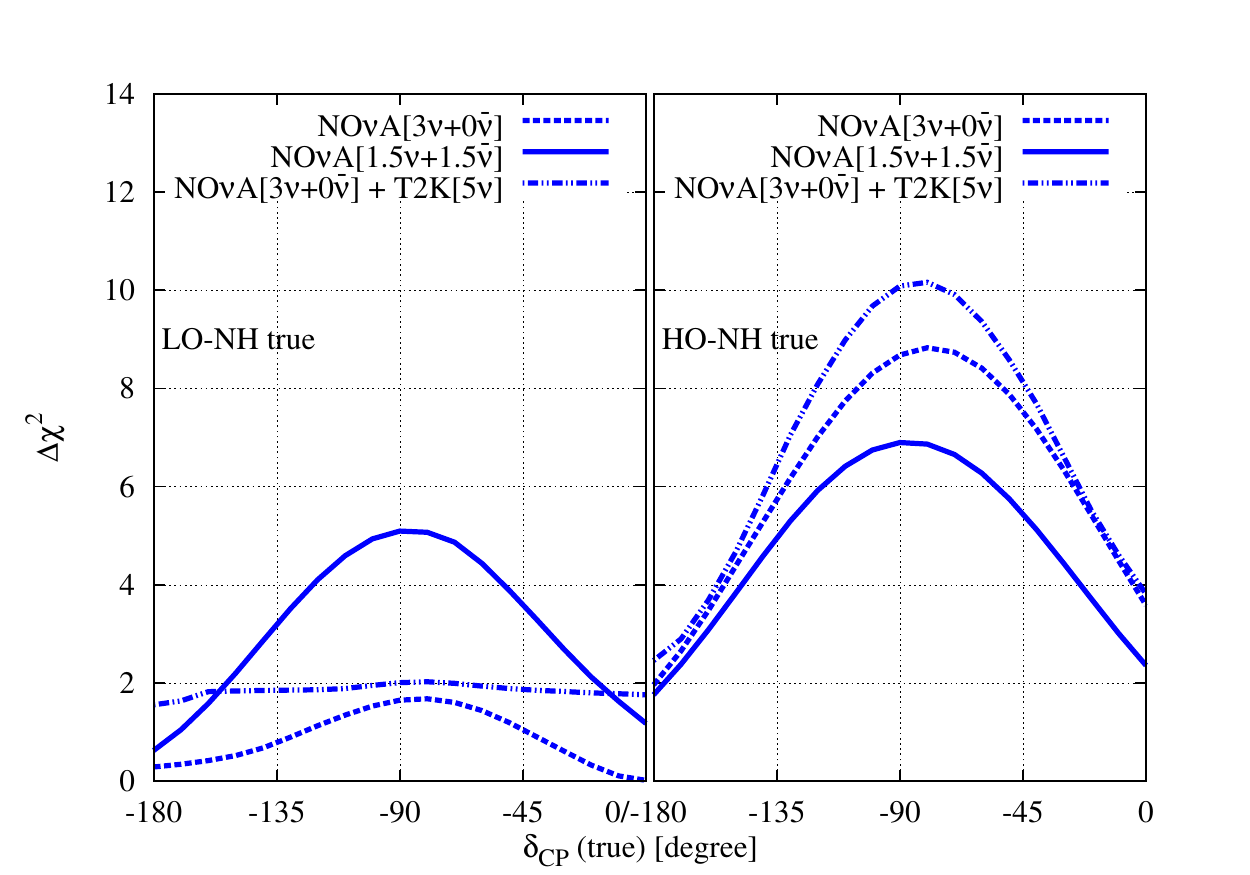}
\caption{\footnotesize{Hierarchy sensitivity assuming 5$\%$ uncertainty in 
$\sin^22\theta_{13}$ for NH and LHP. In the left (right) panel, 
the true $\sin^2\tz$ is taken to be 0.45 (0.59).}}
\label{newtzNH}
\end{figure}

\begin{figure}[H]
\centering
\includegraphics[width=0.8\textwidth]{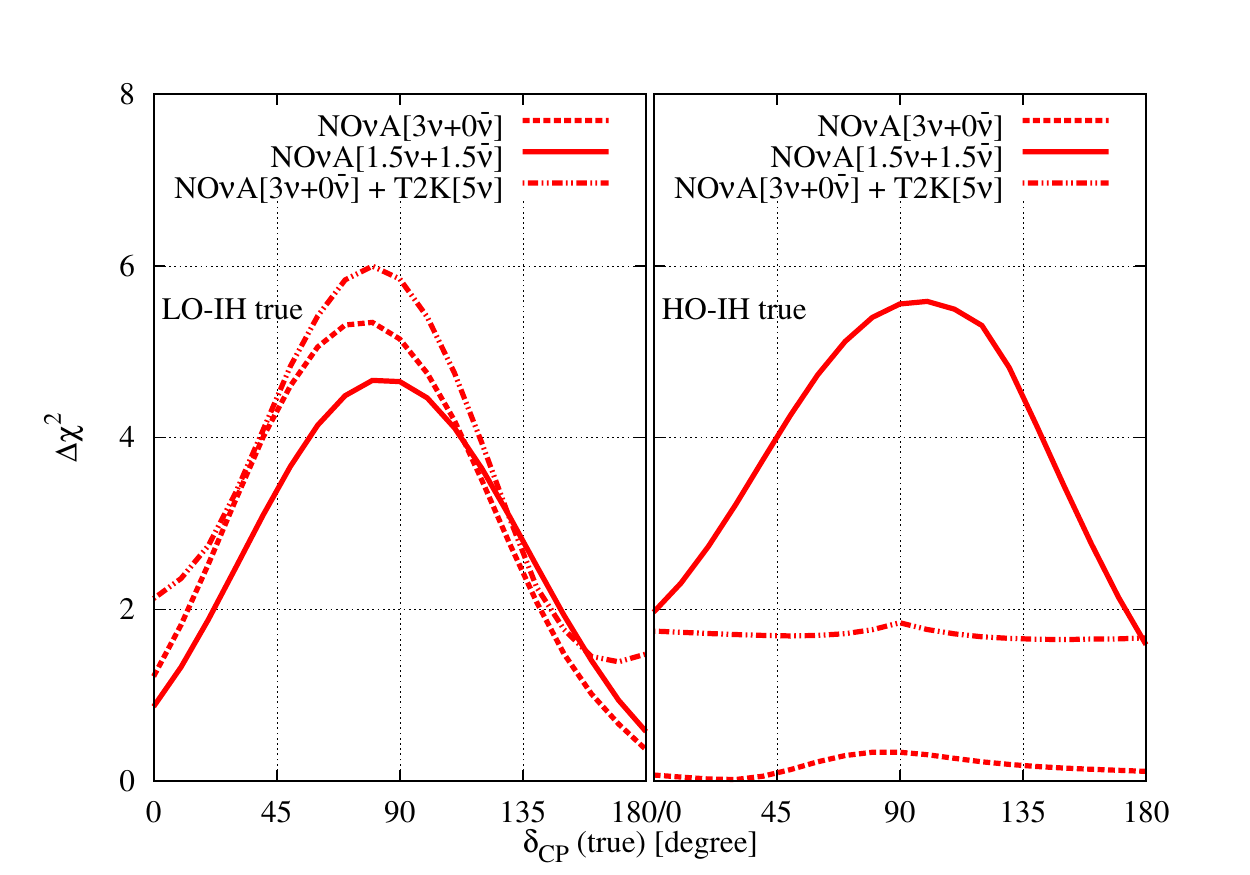}
\caption{\footnotesize{Hierarchy sensitivity assuming 5$\%$ uncertainty in 
$\sin^22\theta_{13}$ for IH and UHP. In the left (right) panel, 
the true $\sin^2\tz$ is taken to be 0.45 (0.59).}}
\label{newtzIH}
\end{figure}

The most recent results of the T2K experiment \cite{Abe:2014ugx}
give $\sin^2 \tz = 0.514^{+0.055}_{-0.055} (0.511 \pm 0.055)$
for NH (IH). These values seem to favour maximal mixing
but a deviation from maximality is also very likely. The 
parameters we have chosen here fall within the $2 \sigma$ 
range of these measurements. Even if the deviation of $\tz$
from maximality is very small ($|\sin^2 \tz - 0.5| = 0.02$),
the hierarchy sensitivity of $1.5 \nu + 1.5 \anu$ run is better
than that of $3 \nu$ run for the two combinations LO-NH and HO-IH.
This is illustrated in figs.~(\ref{smalldevNH}) and~(\ref{smalldevIH}).

\begin{figure}[H]
\centering
\includegraphics[width=0.8\textwidth]{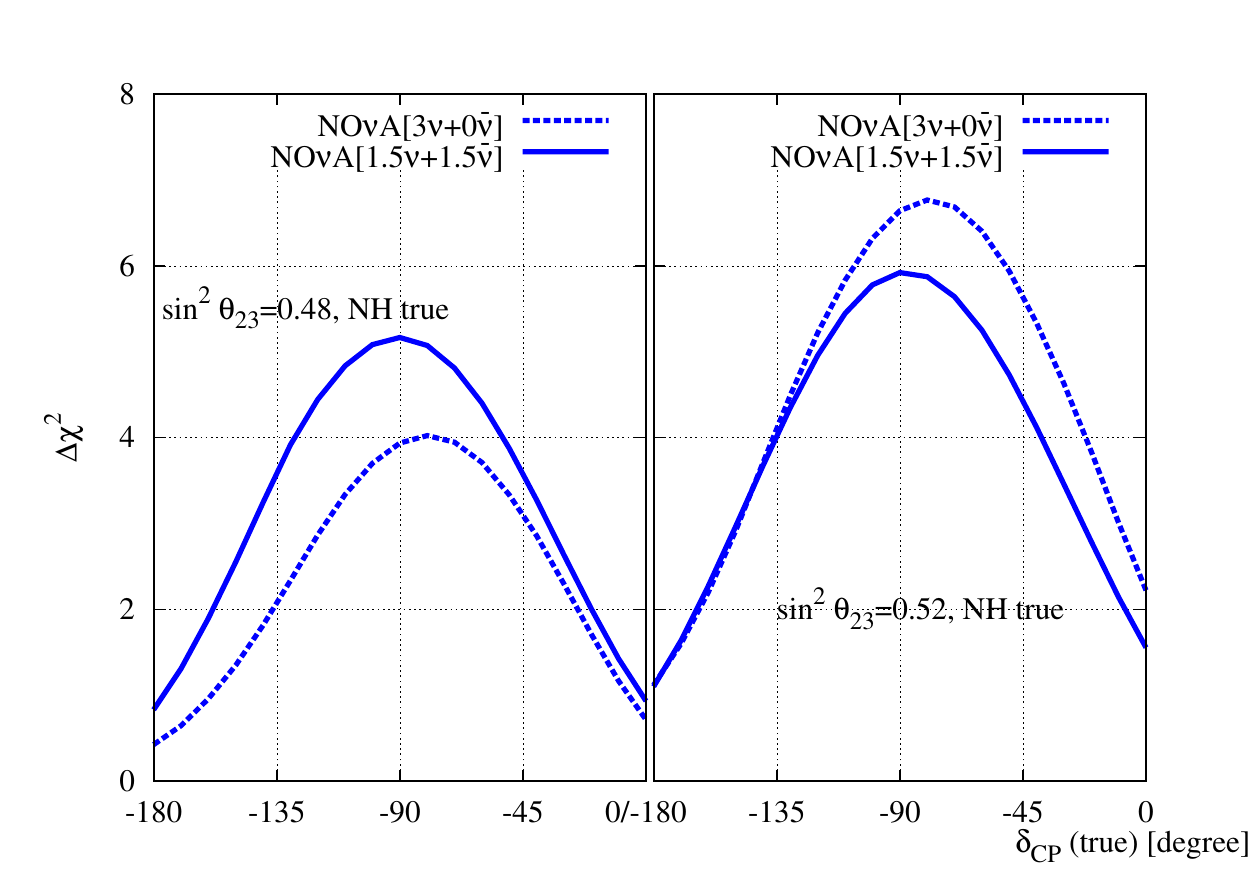}
\caption{\footnotesize{Hierarchy sensitivity assuming 5$\%$ uncertainty in 
$\sin^22\theta_{13}$ for NH and LHP. In the left (right) panel, 
the true $\sin^2\tz$ is taken to be 0.48 (0.52).}}
\label{smalldevNH}
\end{figure}

\begin{figure}[H]
\centering
\includegraphics[width=0.8\textwidth]{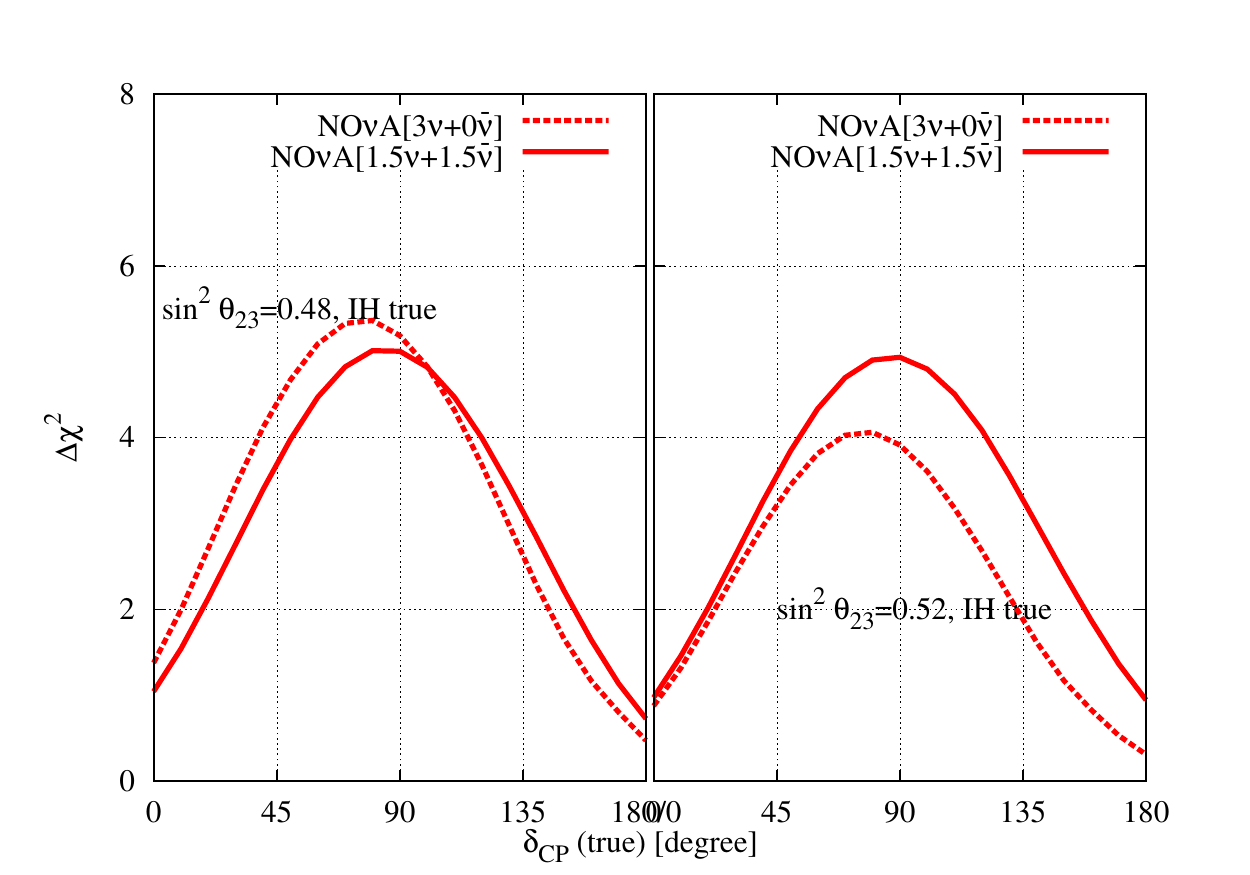}
\caption{\footnotesize{Hierarchy sensitivity assuming 5$\%$ uncertainty in 
$\sin^22\theta_{13}$ for IH and UHP. In the left (right) panel, 
the true $\sin^2\tz$ is taken to be 0.48 (0.52).}}
\label{smalldevIH}
\end{figure}

\subsection{Advantages of $1.5\nu+1.5\anu$ run of \nova}

In the previous subsection, we have argued that the $1.5 \nu + 1.5 \anu$
run of \nova has good hierarchy sensitivity for all four combinations of 
octant and hierarchy whereas the $3 \nu$ run has a slightly better hierarchy
sensitivity for the two combinations HO-NH and LO-IH. Thus it becomes 
important to address
the question: Can 1.5 years of $\nu$ data of \nova give a hint of hierarchy
if either HO-NH or LO-IH are the true combinations? 
Based on the results of the previous sub-section, we know that there 
will be no sensitivity if LO-NH or HO-IH are true. For the other two
cases, HO-NH and LO-IH, the hierarchy sensitivity from the $1.5 \nu$
data is given in Fig.~\ref{1.5run}. From this figure, we see that 
there is reasonable hierarchy sensitivity for the combination HO-NH, 
even from 1.5 years of $\nu$ data, but not for the combination LO-IH.
This is expected because $\pmue$ receives a double boost in the case
of HO-NH and hence there will be a large number of signal events.
For LO-IH, $\pmue$ gets a double suppression and hence the statistics
in the $1.5 \nu$ run are not sufficient to rule out the wrong hierarchy.
Addition of 2 years of $\nu$ data from T2K leads to no significant change.

\begin{figure}[H]
\centering
\includegraphics[width=0.8\textwidth]{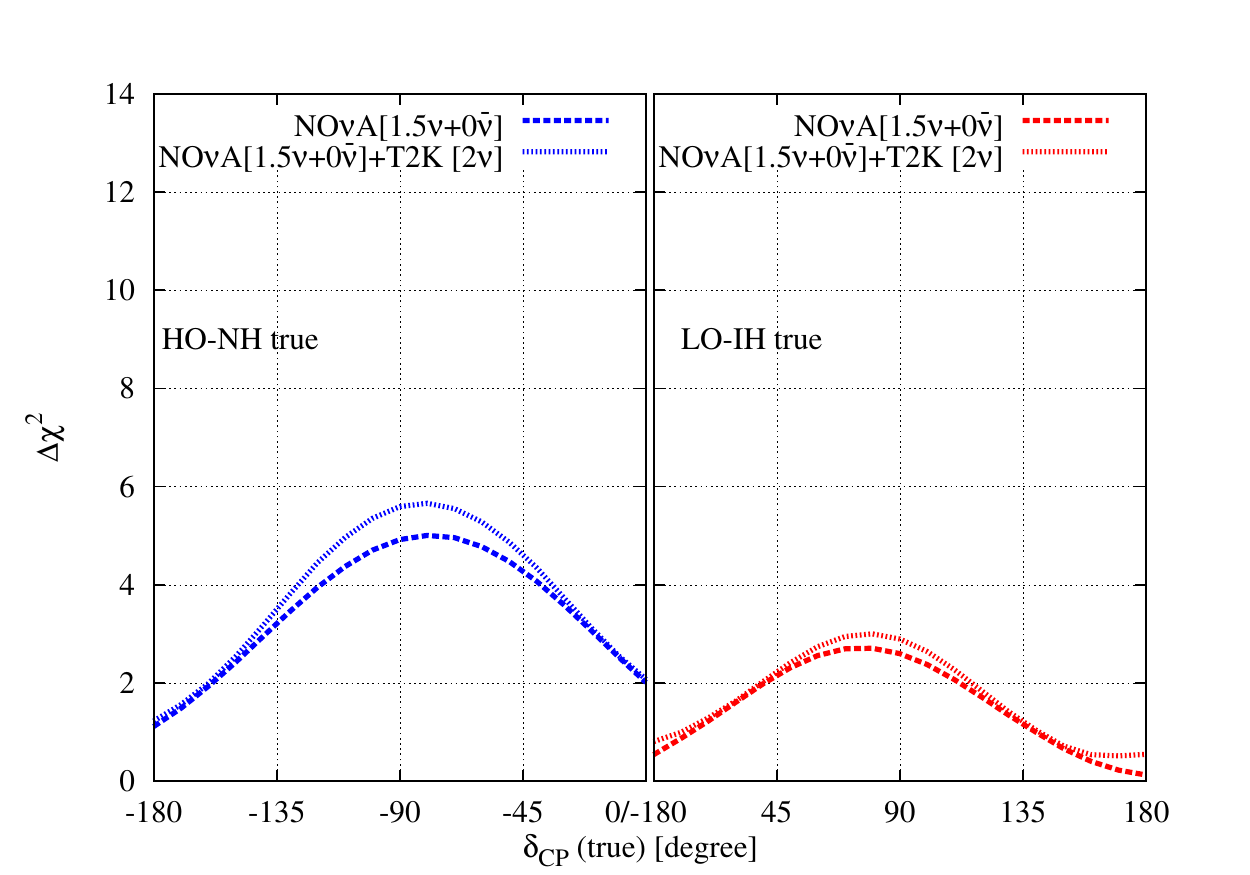}
\caption{\footnotesize{Hierarchy sensitivity of 1.5 years of $\nu$ run  
for HO-NH (left panel) and LO-IH (right panel). In the left (right) panel, 
the true $\sin^2\tz$ is taken to be 0.59 (0.41).}}
\label{1.5run}
\end{figure}

This leads us to a very interesting conclusion: The physics
capabilities of \nova are enhanced if it has $1.5\nu+1.5\anu$
runs during the first three years. This statement is true for 
any octant-hierarchy combination. 
We see above that, for the combination of HO-NH, a $2 \sigma$ hint of hierarchy
is possible for half of LHP, even with 1.5 years of $\nu$ run. If the
hierarchy is known after such a run, then a run plan, which has the best 
CP sensitivity, is preferable. To maximize the CP sensitivity,
it is desirable to have equal number of $\nu$ and $\anu$ events
\cite{Campagne:2006yx}. This requires a longer $\anu$
run because the $\anu$ cross sections are smaller. Hence, 
if HO-NH is true, a hierarchy hint can be obtained with a $1.5 \nu$ run, 
after which it is preferable to run \nova in $\anu$ mode only.  
For the other three octant-hierarchy combinations, $1.5\nu$ run does
not give a hint of hierarchy. In such a situation, a switch to
$\anu$ run will guarantee a $2 \sigma$ hierarchy discrimination for
a reasonable fraction of the favourable half plane of $\dcp$.

\section{Conclusions}

\nova experiment is about to start taking data. Among its physics
goals are (a) the determination of neutrino mass hierarchy, (b) 
the determination of the octant of $\tz$ and (c) the discovery of leptonic
CP violation. 
The hierarchy reach of pure 
$\nu$ data is subject to $\ty$-hierarchy and octant-hierarchy 
degeneracies, whereas equal $\nu$-$\anu$ runs are free from them.
If the uncertainty in $\sin^2 2 \ty$ remains at the present $10\%$
level, then the combination $1.5 \nu + 1.5 \anu$ run has better 
hierarchy sensitivity compared to pure $3 \nu$ run.
Even when this uncertainty is reduced to $5\%$, the $3\nu$
run {\bf fails} to give any hierarchy discrimination, if
the true combinations are LO-NH or HO-IH, whereas the combined
$1.5 \nu + 1.5 \anu$ run has good hierarchy discrimination for
all four octant-hierarchy combinations. 

We argue that it is advantageous for \nova to have equal
1.5 years of $\nu$ and $\anu$ runs during the first three years.
We find that $1.5\nu$ run gives a $2 \sigma$ hierarchy hint if 
the combination HO-NH is true and $\dcp$ is in LHP.
In such a situation, it is better to switch to $\anu$ to maximize
the CP sensitivity. For the other three octant-hierarchy combinations, $1.5 \nu$ 
run has poor or no hierarchy sensitivity. Following this up with a
1.5 year $\anu$ run will give a better chance of hierarchy discrimination, if $\dcp$
is in the favourable half plane.

Finally, what should happen after $1.5 \nu + 1.5 \anu$ run? If no hint of hierarchy is 
obtained, then a farther $1.5\nu+1.5\anu$ run seems preferable.
Then, the full hierarchy discrimination capability of $3\nu+3\anu$ run 
of \nova will be realised. If a hint of hierarchy is found, then having
the additional run in $\anu$ mode is likely to give the best CP sensitivity.

\underline{\it Acknowledgements}: 
We thank Sanjib Agarwalla for discussions on implementing
re-tuned \nova efficiencies in GLoBES. We thank Maury Goodman, Mark Messier 
and Jon Urheim for their comments on the manuscript. S. U. S. thanks
Stephen Parke, Peter Shanahan and other members of \nova group at Fermilab and 
Carlos Wagner, Maury Goodman and other members of \nova group at Argonne for their
hospitality and for discussions related to this paper. We thank Srubabati 
Goswami for a critical reading of the manuscript.
U. R. thanks Council for Scientific and Industrial Research (CSIR), 
Government of India, for financial support.

\bibliographystyle{apsrev}
\bibliography{referenceslist}

\end{document}